\documentclass[aps,prc,twocolumn,groupedaddress,showpacs]{revtex4-1}
\usepackage{color}
\usepackage{graphicx}
\usepackage{hyperref}

\newcommand{\jpsi}{\mathrm{J/}\psi}

\newcommand{\WgPb}{W_{\gamma\mathrm{Pb}}}
\newcommand{\Wgp}{W_{\gamma\mathrm{p}}}
\newcommand{\sNN}{\sqrt{s_{\rm NN}}}
\newcommand{\gPb}{\gamma\mathrm{Pb}}
\newcommand{\gp}{\gamma\mathrm{p}}

\begin{document}

%\preprint{}

%\title{\boldmath 
%Small $x$ gluon shadowing from LHC data on coherent  $\mathrm{J/}\psi$ photoproduction
%\unboldmath}

\title{\boldmath 
Gluon shadowing at small $x$ from coherent $\mathrm{J/}\psi$ photoproduction data
at energies available at the CERN Large Hadron Collider
\unboldmath}


\author{J. G. Contreras}
%\thanks{}
\affiliation{Faculty of Nuclear Sciences and Physical Engineering,
Czech Technical University in Prague, Czech Republic}


\date{\today}

\begin{abstract}
The cross section for coherent $\jpsi$ photoproduction in Pb-Pb collisions  is the sum of two contributions, one from low-, the other from high-energy photon-nucleus interactions.
A novel method to disentangle both contributions allowing one to extract the coherent  photo-nuclear cross section for coherent $\jpsi$ production,  $\sigma_{\gamma \rm Pb}$, is presented. 
The utility of the method is demonstrated using 
 measurements from peripheral and ultra-peripheral Pb-Pb collisions at the LHC. Applying the proposed method to the available data  it is possible to obtain  
$\sigma_{\gamma \rm Pb}$ up to a photon-lead centre-of-mass energy of  470 GeV, which corresponds to $x$ of 4.4x10$^{-5}$. To illustrate a possible use of the extracted photo-nuclear cross sections, the corresponding nuclear suppression factors are computed and  they are  compared to  predictions of gluon shadowing calculated in the leading-twist approximation approach.
\end{abstract}

% 13.60.-r 	Photon and charged-lepton interactions with hadrons 
% 13.60.Le 	Meson production
% 14.40.Pq 	Heavy quarkonia
% 12.38.-t 	Quantum chromodynamics (for quarks, gluons, and QCD in nuclear reactions, see 24.85.+p)
% 24.85.+p 	Quarks, gluons, and QCD in nuclear reactions
% 25.20.-x 	Photonuclear reactions
% 25.20.Lj 	Photoproduction reactions
%\keywords{}

\maketitle


\section{Introduction}

Shadowing refers to the phenomenon that, for small values of $x$, the distribution of a parton in a nucleon bounded inside a nucleus is suppressed with respect to that of a parton in a free nucleon~\cite{Armesto:2006ph}.
As gluons are the dominant component of nucleons in
 this kinematic domain~\cite{Abramowicz:2015mha},  the case of gluon shadowing is particularly important and has attracted the attention of theorists since a long time; e.g.,~\cite{Mueller:1985wy,Nikolaev:1990ja}.

An experimental observable that is well suited to study gluon shadowing  is coherent $\jpsi$ photoproduction, because in this case the interaction proceeds by the exchange of at least two gluons~\cite{Ryskin:1992ui,Guzey:2016piu}. New data for this process have recently been made available by  ALICE, with the $\jpsi$ measured   at forward~\cite{Abelev:2012ba} and at mid rapidities~\cite{Abbas:2013oua}.
 CMS has also released  measurements where the $\jpsi$ is produced at semi-forward rapidities~\cite{Khachatryan:2016qhq}. All these results were obtained  in Pb-Pb ultra-peripheral collisions (UPC) during  Run 1 at the 
 LHC.
 
 Surprisingly, ALICE discovered an excess in the production of  $\jpsi$ in peripheral Pb-Pb 
 collisions~\cite{Adam:2015gba}. After discarding all other potential explanations, and given that the excess is located at very low values of the transverse momentum of the $\jpsi$, the authors of~\cite{Adam:2015gba} proposed the "{\it plausible assumption}" that this excess was caused by coherent $\jpsi$ photoproduction in peripheral Pb-Pb collisions and measured the  cross section for this process at forward rapidities. In this work, it will be considered that this "{\it plausible assumption}" is indeed correct.
 
\begin{figure*}
\includegraphics[width=0.48\textwidth]{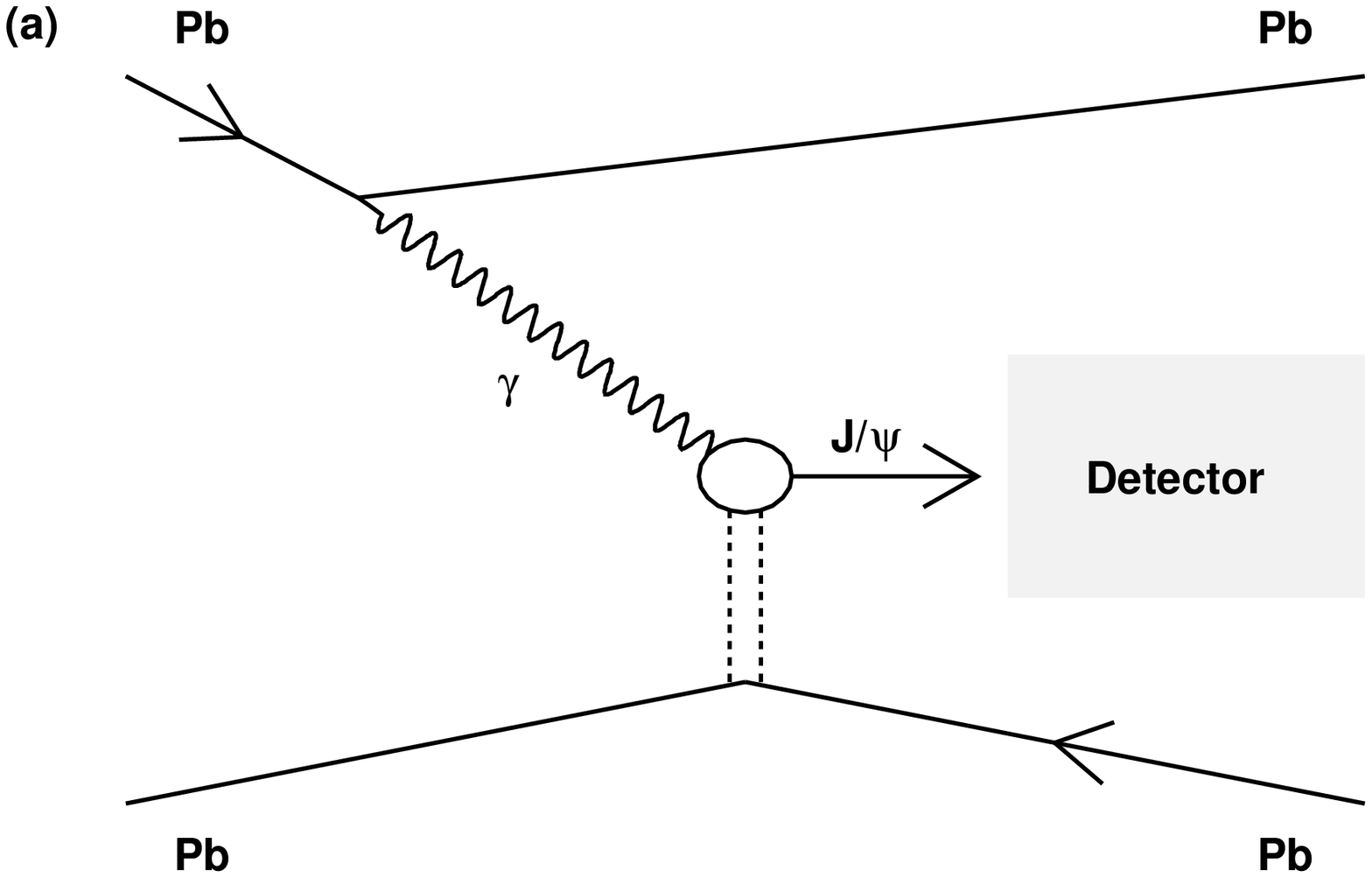}%
\includegraphics[width=0.48\textwidth]{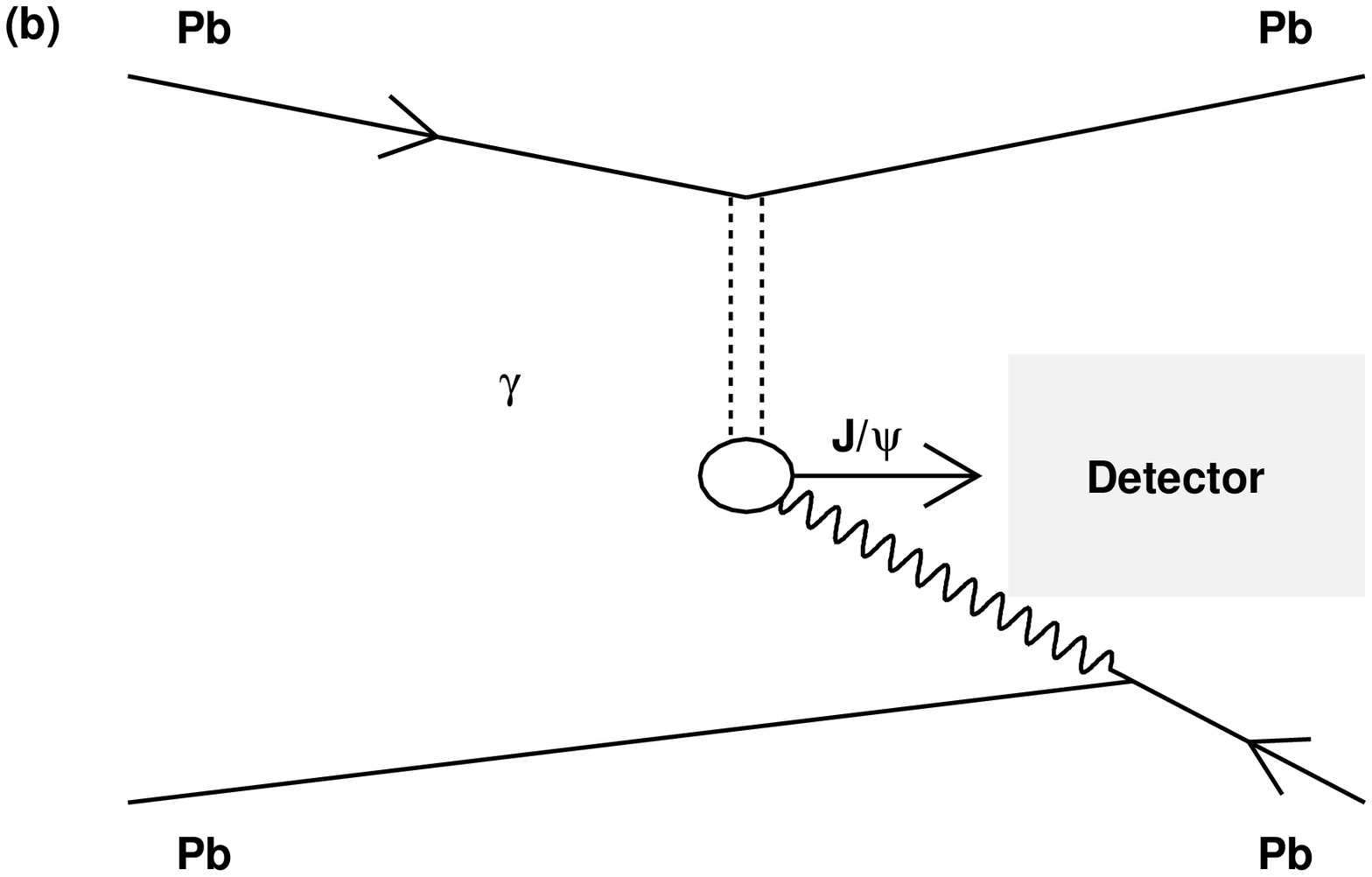}
\caption{\label{fig:diag} Contributions to coherent production of $\jpsi$ in Pb-Pb collisions. In diagram (a) the target lead nucleus is traveling away from the detector at the moment of the interaction, while in diagram (b) it is traveling towards it.}
\end{figure*}

Coherent production of $\jpsi$ in Pb-Pb collisions is a two-step process where first one of the lead nucleus (denoted in the following as source) emits a quasi-real photon which interacts with the second lead nucleus (denoted in the following as target) to coherently produce the $\jpsi$. As such, this process has two contributions as illustrated in Fig.~\ref{fig:diag}. In diagram (a), at the moment of the interaction, the source (target) nucleus is traveling towards (away from) the detector while in diagram (b) it is the other way around.

Accordingly, the cross section $d\sigma_{\rm PbPb}/dy$ for the coherent photoproduction of a $\jpsi$ at rapidity $y$ in Pb-Pb collisions can be factorized as the product of the photon flux $n_\gamma(y,\{b\})$ and the  photo-nuclear  cross section $\sigma_{\gamma \rm Pb}(y)$ as:
  
\begin{equation}
\frac{d\sigma_{\rm PbPb}}{dy} = 
n_{\gamma}(y;\{b\})\sigma_{\gamma \rm Pb}(y)+
n_{\gamma}(-y;\{b\})\sigma_{\gamma \rm Pb}(-y), 
\label{eq:XS}
\end{equation}
where  $\{b\}$ delimits the impact-parameter range taken into account in the measurement. 

The rapidity $y$ in Eq. (\ref{eq:XS}) is given in the laboratory frame. It is defined with respect to the direction of the target.  The rapidity of the $\jpsi$ is related to the center-of-mass energy of the photon-lead system through 
\begin{equation}
\WgPb^2 = \sqrt{s_{\rm NN}}M_{\jpsi}e^{-y},
\end{equation}
where $M_{\jpsi}$ is the mass of the $\jpsi$ vector meson and $\sqrt{s_{\rm NN}}$  is the center-of-mass energy per nucleon pair in the Pb-Pb system. During  Run 1 at the LHC, when ALICE measurements were performed, $\sqrt{s_{\rm NN}}$ was 2.76 TeV. Measurements to be performed at the LHC during Run~2  will have $\sqrt{s_{\rm NN}}=5.02$ TeV. 

In a theoretical description of this process, the emission of the quasi-real photon in Fig.~\ref{fig:diag} and  in Eq. (\ref{eq:XS}) is described by standard electromagnetism, while the interaction of the photon with the target involves the exchange of gluons and it is computed within QCD. Coherent production in the case of UPC  has been studied by several authors (for a recent review including LHC Run1 data see ~\cite{Contreras:2015dqa} and references therein), while  given its recent and unexpected character, production in peripheral collisions has been studied up-to-now only in~\cite{Klusek-Gawenda:2015hja}.

Note that both contributions in Eq. (\ref{eq:XS}) are equal for a measurement of the $\jpsi$ at mid-rapidity, $y=0$, while for measurements at other rapidities the photo-nuclear interaction in  diagram (a) occurs at a larger $\WgPb$ than in diagram (b). The measurement at a given rapidity is the sum of both contributions; that is, it corresponds to the left-hand side of Eq. (\ref{eq:XS}), while of interest for QCD is the photo-nuclear cross section $\sigma_{\gamma \rm Pb}(y)$.

In this work  a novel method to extract  $\sigma_{\gamma \rm Pb}$ is presented. The only ingredient of the method is the computation of the photon fluxes, which is carried out within standard electromagnetism.  With these fluxes the cross section  $\sigma_{\gamma \rm Pb}$ can be extracted using Eqs.~(\ref{eq:SigYneg}) and~(\ref{eq:SigYpos}), which are the main result of this article. As an example of their utility, these equations are then applied to  currently available ALICE data on  coherent production of $\jpsi$ in peripheral and ultra-peripheral Pb-Pb collisions  to extract $\sigma_{\gamma \rm Pb}$ up to a photon-lead center-of-mass energy of  470 GeV.
Finally, as a further illustration of the usefulness of this method and in order to study nuclear shadowing in an standard way, the nuclear suppression factor $S_{\rm Pb}(\WgPb)$ is compared to a current prediction of gluon shadowing computed in the leading-twist approximation~\cite{Guzey:2013qza}.

The rest of the article is organized as follows. Next section describes the computation of the photon fluxes. In Sec.~\ref{sec:extraction} the formulas to extract the photo-nuclear cross section $\sigma_{\gamma \rm Pb}$ are presented and applied to ALICE data to obtain $\sigma_{\gamma \rm Pb}$ at three different values of $\WgPb$. The nuclear suppression factor $S_{\rm Pb}(\WgPb)$ is introduced in Sec.~\ref{sec:nsf} where it is compared to the gluon shadowing predictions from~\cite{Guzey:2013qza}. The results presented here are discussed in Sec.~\ref{sec:discussion} and the article is closed with a brief summary and outlook in Sec.~\ref{sec:so}.

\section{The photon flux}

\subsection{Photon flux and form factors}
For a relativistic charged particle with Lorentz factor $\gamma$, the electromagnetic field is dominated by the  component transverse to the direction of movement of the particle. In this case the field can be interpreted as a flux of quasi-real photons according to an idea proposed by Fermi~\cite{Fermi:1924tc,Fermi:1925fq} and later refined by Weizs\"acker~\cite{vonWeizsacker:1934sx} and Williams~\cite{Williams:1934ad}. For heavy ions this photon flux can be reliably computed in the semiclassical description, see e.g.~\cite{Krauss:1997vr,Baur:2001jj}.

In this formalism the flux of quasi-real photons emitted  with an energy $k$ at a distance $\vec{b}$ from the center of the charged particle is given by:

\begin{equation}
n(k,\vec{b}) = \frac{Z^2\alpha}{\pi^2k}
\left|\int_0^\infty dk_{\perp} k^2_\perp \frac{F(k^2_\perp+(k/\gamma)^2}{k^2_\perp+(k/\gamma)^2}
J_1(b k_\perp)
\right|^2,
\label{eq:FluxWithFF}
\end{equation}
where $Z$ is the electric charge of the particle, $\alpha$ is the QED fine structure constant, $J_1$ a Bessel function, $b$ and $k_\perp$ the magnitudes of  $\vec{b}$ and $\vec{k}_\perp$, respectively, and $F$ is the form factor of the charged particle.

Several prescriptions have been considered for the form factor to be used to describe coherent photoproduction of $\jpsi$ in Pb-Pb collisions. They are presented and compared in the following. The easiest form factor is that of  a point charge ($pc$), $F_{pc}(q) = 1$. In this case,  the integral present in Eq.~(\ref{eq:FluxWithFF}) can be performed analytically yielding
\begin{equation}
n_{pc}(k,\vec{b}) = \frac{Z^2\alpha_{\rm QED}k}{\pi^2\gamma^2}
K^2_1(kb/\gamma).
\end{equation}

To have a more realistic description of a lead nucleus, it has been considered to model the form factor as  a convolution of a hard sphere with radius $R_A$ and a Yukawa potential with range $a$ \cite{Davies:1976zzb} given by

\begin{equation}
F_{hsY}(q) = \frac{4\pi d_0}{Aq^3}\left[\sin(qR_A)-qR_A\cos(qR_A)\right]\left(\frac{1}{1+a^2q^2}\right),
\label{eq:FhsY}
\end{equation}
where for the case of Pb the following values for the parameters have been used: $d_0 = 0.13815$ fm$^{-3}$, $R_A = 1.2A^{1/3}$ fm and $a= 0.7$ fm. 

Another way to describe the form factor of lead nuclei is to start with a  Woods-Saxon nuclear density profile:
\begin{equation}
\rho(b) = \frac{\rho_0}{1+\exp(\frac{b-r_A}{z_0})},
\label{eq:WSrho}
\end{equation}
where for the case of Pb the following values have been used: $\rho_0=0.16$ fm$^{-3}$, $r_A=6.624$ fm and $z_0=0.549$ fm. The form factor is obtained from this density profile through a
 Fourier-Bessel transform:
 
 \begin{equation}
F_{WS}(q) = \frac{4\pi}{qA}\int \rho(b)\sin(bq)bdb.
\label{eq:FWS}
\end{equation}

\begin{figure}[!t]
\includegraphics[width=0.48\textwidth]{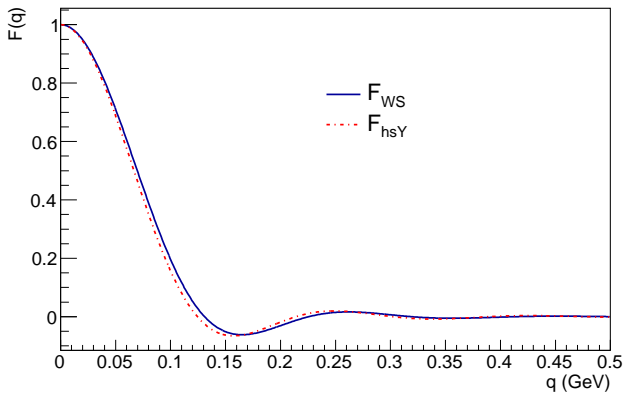}
\caption{
\label{fig:FF} 
Form factor for lead computed using Eq.~(\ref{eq:FhsY}) for the dash-dot red line and Eq.~(\ref{eq:FWS}) for the solid blue line.
}
\end{figure}

Figure \ref{fig:FF} compares the form factor $F_{hsY}$ with $F_{WS}$, where this last one has been obtained by numerical integration of Eq.~(\ref{eq:FWS}). Given that both form factors are very similar, and to avoid the numerical integration in $F_{WS}$, in the following $F_{hsY}$ has been used.

\begin{figure}
\includegraphics[width=0.48\textwidth]{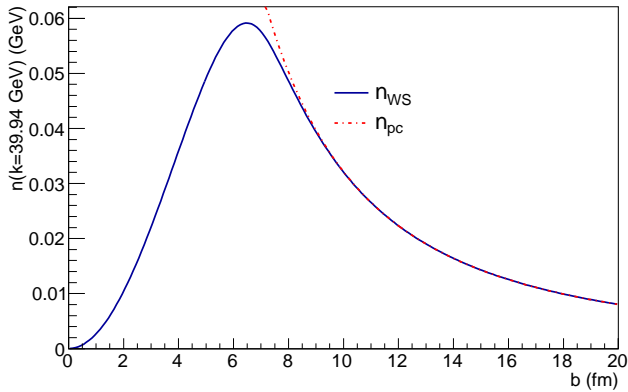}
\caption{
\label{fig:Nkx} 
Photon flux according to Eq.~(\ref{eq:FluxWithFF}) for a photon energy of $k=39.94$ GeV (see Table~\ref{tab:extrainfo}) using the form factor of Eq.~(\ref{eq:FhsY}) for the solid blue line or the point charge form factor for the dash-dot red line.
}
\end{figure}

Figure \ref{fig:Nkx} compares the photon fluxes computed with  the form factors of a point charge $F_{pc}$ and  $F_{hsY}$ given by Eq.~(\ref{eq:FhsY}). The comparison is performed at a photon energy relevant for the discussion below, but a similar picture emerges when using other photon energies. The figure shows that both fluxes agree quite well down to 8-9 fm.  

\subsection{Probability of no hadronic interaction}

The experimental requirements imposed in the measurements performed by ALICE restrict the impact parameter range sampled in peripheral and ultra-peripheral interactions. Following~\cite{Klein:1999qj}, the probability $P_{NH}$ of having no hadronic interaction between the incoming lead nuclei is used to take into account the experimental situation.
This probability is computed starting  from the 
 Woods-Saxon distribution given in Eq.~(\ref{eq:WSrho}). Then the nuclear thickness function is defined as

 \begin{equation}
T_A(\vec{b}) = \int dz\rho(\sqrt{|\vec{b}|^2+z^2}), 
\end{equation}
where $z$ is the direction of travel of the nucleus, which is perpendicular to the impact parameter plane where $\vec{b}$ is defined. Then the nuclear overlap function is defined by
 \begin{equation}
T_{AA}(|\vec{b}|) = \int d^2\vec{r}\ T_A(\vec{r})T_A(\vec{r}-\vec{b}).
\end{equation}

The probability of a nucleon-nucleon interaction in the collision of two nuclei can be seen  as a Poisson process with mean $T_{AA}\sigma_{NN}$ where $\sigma_{NN}$ is the  nucleon-nucleon inelastic cross section. In this case the probability of no hadronic interaction in the collision of two nuclei at an impact parameter $b$ is given by

 \begin{equation}
P_{NH}(b) =\exp(-T_{AA}\sigma_{NN}).
\label{eq:PNH}
\end{equation}

\begin{figure}
\includegraphics[width=0.48\textwidth]{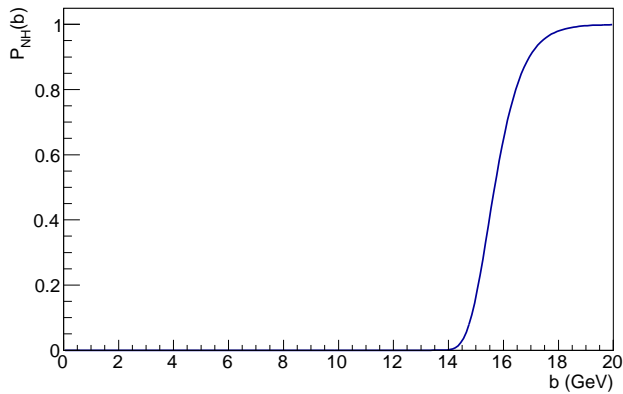}
\caption{
\label{fig:PNH} 
Probability of no hadronic nuclear interaction according to Eq.~(\ref{eq:PNH}).
}
\end{figure}

For the computations  $\sigma_{NN} = 64$ mb has been used. The probability of no hadronic interaction is shown in Fig. \ref{fig:PNH}. It can be seen that this probability is basically zero below 14 fm.

\subsection{The photon fluxes for peripheral and ultra-peripheral collisions}

The experimental requirement in UPC is that the nuclei remain intact after the interaction that produces the $\jpsi$. The photon flux implementing this requirement  is given by

 \begin{eqnarray}
n^U_\gamma(y) &=& k\int_0^\infty db 2\pi b P_{NH}(b) \nonumber \\ 
& & \int_0^{r_A}\frac{rdr}{\pi r^2_A}\int_0^{2\pi}d\phi n(k,b+r\cos(\phi)).
\label{eq:nU}
\end{eqnarray}
The superscript $U$ represents the impact parameter range taken into account in this case. This was represented in general by $\{b\}$ in Eq~(\ref{eq:XS}). The factor of $k$ in front of the integral comes from transforming rapidity of the $\jpsi$ to photon energies using 
\begin{equation}
k = \frac{M}{2}e^y.
\end{equation}

The second line of Eq.~(\ref{eq:nU}) performs an average over the surface of the {\it target} nucleus. Such a factor was introduced in~\cite{Klein:1999qj}  to take into account the coherent condition.

The experimental requirement for peripheral collisions is that the nuclei undergo a hadronic collision. In this case the corresponding flux reads
 \begin{eqnarray}
n^P_\gamma(y) &=& k\int_{b_{\rm min}}^{b_{\rm max}} db 2\pi b (1-P_{NH}(b)) \nonumber \\ 
& & \int_0^{r_A}\frac{rdr}{\pi r^2_A}\int_0^{2\pi}d\phi n(k,b+r\cos(\phi)).
\label{eq:nP}
\end{eqnarray}

In this case the superscript $P$ represents the impact parameter range $(b_{\rm min},b_{\rm max})$. The measurement reported in~\cite{Adam:2015gba}  was performed in the centrality class 70\%-90\%, which according to the centrality determination by ALICE corresponds to an impact parameter range from $b_{\rm min} = 13.05$ fm to $b_{\rm max} = 14.96$ fm~\cite{Abelev:2013qoq}. 

\section{Extraction of the coherent photo-nuclear cross section\label{sec:extraction}}

\subsection{The method}
As it has been shown above, the computation of the photon fluxes entering Eq.~(\ref{eq:XS}) can be performed using standard electromagnetism. 
This opens up the possibility that
 the comparison of two Pb-Pb cross sections at the same rapidity, but in different regions of  impact parameter, allows one to extract the coherent photo-nuclear cross section both at $y$ and at $-y$. 
 
 Indeed, using the superscripts $P$ and $U$ for convenience to denote the two cross sections and solving the  set of equations obtained from using Eq.~(\ref{eq:XS}) for each one of them, it is obtained that:
\begin{equation}
\sigma_{\gamma {\rm Pb}}(-y) =
\left(n^P_{\gamma}(y)\frac{d\sigma^U_{\rm PbPb}}{dy} -n^U_{\gamma}(y)\frac{d\sigma^P_{\rm PbPb}}{dy} \right)/F(y),
\label{eq:SigYneg}
\end{equation}
and
\begin{equation}
\sigma_{\gamma {\rm Pb}}(y) =
\left(n^U_{\gamma}(-y)\frac{d\sigma^P_{\rm PbPb}}{dy} -n^P_{\gamma}(-y)\frac{d\sigma^U_{\rm PbPb}}{dy} \right)/F(y)
\label{eq:SigYpos}
\end{equation}
where
\begin{equation}
F(y)\equiv
n^P_{\gamma}(y)n^U_{\gamma}(-y)
-n^U_{\gamma}(y)n^P_{\gamma}(-y).
\label{eq:Flux}
\end{equation}
In here, $P$ and $U$ are two arbitrary cross sections which are measured at the same rapidity, but in a different range of impact parameters. 

\subsection{Existing data}

Currently, there are data from ALICE, which are not ideal, but that can be used to perform a first extraction of the photo-nuclear cross section using Eqs.~(\ref{eq:XS}), (\ref{eq:SigYpos}) and~(\ref{eq:SigYneg}).  

The cross section for  coherent production of $\jpsi$ in Pb-Pb collisions at mid rapidity, as reported in~\cite{Abbas:2013oua} is
\begin{equation}
\frac{d\sigma^{U}_{\rm PbPb}}{dy} (|y|<0.9) =  
2.38^{+0.34}_{-0.24}\ {\rm (stat.+syst.)} \ {\rm mb}.
\label{eq:XSmidUPC}
\end{equation}
In this case, Eq.~(\ref{eq:XS}) can be applied directly using the photon flux given by Eq.~(\ref{eq:nU}).

Furthermore, there are two measurements where the $\jpsi$ is produced at forward rapidities. One for UPC is reported in \cite{Abelev:2012ba}
\begin{eqnarray}
\frac{d\sigma^{U}_{\rm PbPb}}{dy}&& (2.6<|y|<3.6) = \nonumber \\
& & 1.00\pm0.18\  {\rm (stat.)} \ ^{+0.23}_{-0.26}
\ {\rm (syst.)} \ {\rm mb}, 
\label{eq:XSfwdUPC}
\end{eqnarray}
and one for peripheral collisions reported in~\cite{Adam:2015gba}: 
\begin{eqnarray}
\frac{d\sigma^{P}_{\rm PbPb}}{dy}&& (2.5<|y|<4.0) =  \nonumber \\
& & 59\pm11\  {\rm (stat.)} \ ^{+10.6}_{-12.8}
\ {\rm (syst.)} \ \mu{\rm b}. 
\label{eq:XSfwdP}
\end{eqnarray}

Using the two latter  measurements, Eqs.~(\ref{eq:nU}) and~(\ref{eq:nP}), as well as Eqs.~(\ref{eq:SigYpos}) and~(\ref{eq:SigYneg}) it is possible to perform a first extraction of  $\sigma_{\gamma \rm Pb}$ for two more $\WgPb$ energies.

\subsection{Extraction of $\sigma_{\gamma \rm Pb}$}
As already mentioned these data are not ideal to be used with the method proposed here. There are two main problems. One is that the method works at a fixed value of rapidity and each measurement has been performed over a broad range of rapidities. This raises the issue of which rapidity to chose to apply the method. The second problem is that the measurements~(\ref{eq:XSfwdUPC}) and ~(\ref{eq:XSfwdP}) have been performed in slightly different rapidity ranges.

\begin{table}%[H] add [H] placement to break table across pages
\caption{\label{tab:extrainfo} Numerical values of the different computed quantities that are used in Eqs.~(\ref{eq:XS}), (\ref{eq:SigYneg}-\ref{eq:SigYpos}) and Eqs. (\ref{eq:spb}) to (\ref{eq:phi}).}
\begin{ruledtabular}
\begin{tabular}{lccc}
$y$ & -3.25 & 0 & 3.25 \\
$\WgPb$ (GeV) & 18.2 & 92.4 & 469.5 \\
$x$ & 2.9x10$^{-2}$& 1.1x10$^{-3}$&4.4x10$^{-5}$\\
$k$ (GeV) & 0.06 & 1.55 & 39.94 \\
$n^P_{\gamma}$ & 5.3 & 5.2 & 0.8 \\
$n^U_{\gamma}$ & 168 &66.5 & 0.7 \\
% $\Phi_{\rm Pb}(|t|_{\rm min})$ (GeV$^2)$ & 116.1 & 146.8 & 146.8 \\
% $d\sigma_{\gp}/dt|_{t=0}$ (nb/GeV$^2$) & 81.2 &320.3 & 1178 \\
$ \sigma^{\rm IA}_{\gPb}$ ($\mu$b) & 9.4 & 47.1 & 173.4 \\
% Lines of table here ending with \\
\end{tabular}
\end{ruledtabular}
\end{table}

To address the first problem  two approaches have been considered: ($i$) the center of the rapidity range has been taken as a representative value for the rapidity of the measurement and ($ii$) a model that describes correctly the experimental UPC data has been used to compute the mean value of the rapidity in the given ranges. The mean values found in this way are 3.02 for $2.6<|y|<3.6$ and 3.13 for $2.5<|y|<4.0$. The same mean value has  been used in the peripheral and ultra-peripheral cases. The model used for this is taken from~\cite{Adeluyi:2012ph}; it is based on a LO computation of the process using gluons defined in the collinear limit and shadowing corrections implemented according to the EPS09 parameterization. It has been shown in~\cite{Abbas:2013oua} that this model gives the best description of currently available data.

The same model has been used to address the second problem. The ratio of cross sections corresponding to the UPC case has been evaluated at the central (mean) values of the rapidity ranges in~(\ref{eq:XSfwdUPC}) and ~(\ref{eq:XSfwdP}). The measured cross section quoted in~(\ref{eq:XSfwdUPC})  has been scaled down using the computed ratio to correspond to the central (mean) value of the measurement quoted in~(\ref{eq:XSfwdP}). The ratios obtained with the model from~\cite{Adeluyi:2012ph} are 1.12 and 1.08 for the central and mean rapidity variants, respectively. To estimate a possible model dependence on these ratios other models have been used~\cite{Klein:1999qj,Rebyakova:2011vf,Cisek:2012yt,Goncalves:2011vf}. A brief description of the physics behind these models can be found in~\cite{Contreras:2015dqa}. Note that these model give a worse description of data than the model in~\cite{Adeluyi:2012ph}. The ratios computed with these other models varied at most by 4.5\% from the quoted ratios.  Given that the experimental uncertainty quoted in ~(\ref{eq:XSfwdUPC})  is 
around 30\% (adding in quadrature statistical and systematic uncertainty) the uncertainty in the ratios was disregarded in the following.
 
The systematic uncertainties in~(\ref{eq:XSfwdUPC}) and ~(\ref{eq:XSfwdP}) are certainly partially correlated. These measurements use the same detector, the data were collected at the same time and several of the methods are the same. For example uncertainties on the overall normalization, on the knowledge of the trigger and detector efficiencies and acceptances will be strongly correlated, while the determination of the centrality or the subtraction of  background will not be. A detailed analysis of the correlations can only be performed by ALICE. Here for definiteness the propagation of systematical uncertainty has been performed assuming a correlation of 0.5. It has been checked that choosing other values does not change the results appreciably.

\begin{figure}
\includegraphics[width=0.48\textwidth]{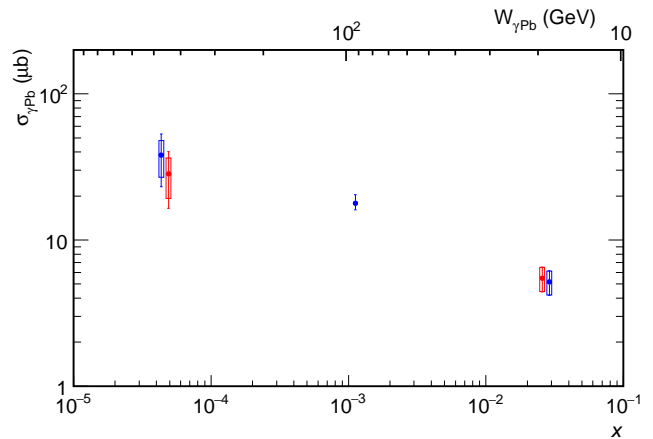}\\%
\caption{\label{fig:XS} Cross section  for $\jpsi$ coherent photo-nuclear production off a lead nucleus as a function of $\WgPb$ (upper axis) and $x$ (lower axis). Statistical uncertainties are represented by a line and systematic uncertainties by the height of the empty box, except for the middle cross section, where the bars represent both types of uncertainties. Blue symbols are evaluated at the center, while red symbols are evaluated at the mean rapidity of the rapidity ranges where the corresponding measurements were performed.}
\end{figure}

For illustration some values used to extract $\sigma_{\gamma \rm Pb}$ are shown in Table~\ref{tab:extrainfo}. 
The  cross sections, defined at the rapidity corresponding to the center of the rapidity range are the following
\begin{eqnarray}
 \sigma_{\gamma {\rm Pb}} &&(\WgPb=18.2\ {\rm GeV}) \nonumber \\
&& = 5.2\pm1.0 \ ({\rm stat.})\ \pm 1.0 \ ({\rm syst.})\ \mu{\rm b},
 \label{eq:XS1} 
 \end{eqnarray}
 
 \begin{eqnarray}
  \sigma_{\gamma {\rm Pb}} & & (\WgPb=92.4\ {\rm GeV})  \nonumber \\
  &&= 17.9^{+2.6}_{-1.8}\ ({\rm stat.+syst.})\ \mu{\rm b},
   \label{eq:XS2}
   \end{eqnarray}
   
\begin{eqnarray}
 \sigma_{\gamma {\rm Pb}}& &(\WgPb=469.5\ {\rm GeV}) \nonumber \\
 && =  38.1\pm 15.0 \ ({\rm stat.})\ ^{+9.9}_{-11.3} \ ({\rm syst.}) \ \mu{\rm b}. 
  \label{eq:XS3}
\end{eqnarray}

Figure~\ref{fig:XS} shows these cross sections as a function of both, $\WgPb$ and $x$, where it has been used that
\begin{equation}
x=\frac{M^2_{\jpsi}}{\WgPb^2}.
\end{equation}
The cross section grows fast with decreasing (increasing) $x$ ($\WgPb$). The point at the smallest $x$ seems to be lower than a simple linear extrapolation, in this log-log scale, from the two points at larger $x$ values. Cross sections extracted using the mean value of the rapidity in the ranges covered by the respective measurements are also shown in this figure. At mid rapidity both approaches produce the same result. Although the numerical values change slightly, the overall picture is the same, irrespectively of the value chosen to represent the rapidity of the measurements.

\section{The nuclear suppression factor\label{sec:nsf}}
\subsection{Definition of the nuclear suppression factor}

In order to highlight the possible effects of nuclear shadowing it is customary to define a so called nuclear suppression factor. This factor compares the cross section of a process in nuclear collisions with the same process in nucleon collisions, properly normalized. In case that there is no shadowing this factor should be one. In case of shadowing it is below one. For the case of coherent production of $\jpsi$ an appropriate nuclear suppression factor has been defined in~\cite{Guzey:2013xba} by:
\begin{equation}
S_{\rm Pb}(\WgPb) =\left( \frac{
\sigma^{\rm data}_{\gPb}(\WgPb)}
{ \sigma^{\rm IA}_{\gPb}(\WgPb)}
\right)^{1/2},
\label{eq:spb}
\end{equation}
where  in the denominator the so-called impulse approximation (IA) is used. The IA is defined as
\begin{equation}
\sigma^{\rm IA}_{\gPb}(\WgPb)  =
\frac{d\sigma_{\gp}(\Wgp=\WgPb,t=0)}{dt}\Phi_{\rm Pb}(|t|_{\rm min}).
\end{equation}
Here, $d\sigma_{\gp}/dt$ at $t=0$ is extracted from data on exclusive $\jpsi$ photoproduction off protons at a photon-proton center-of-mass energy $\Wgp$ according to the fit presented in~\cite{Guzey:2013xba}, $t$ is the square of the  momentum transferred in the target vertex and the  form factor $F_{WS}(t)$ is used to compute
\begin{equation}
\Phi_{\rm Pb}(|t|_{\rm min}) = \int_{|t|_{\rm min}}^\infty d|t|
\left| F_{WS}(t) \right|^2.
\label{eq:phi}
\end{equation}

For illustration, the numerical values for the IA  are reported in Table~\ref{tab:extrainfo}.
Using these values and the results reported in~(\ref{eq:XS1}) to~(\ref{eq:XS3}) the  
 following nuclear suppression factors are obtained:
\begin{eqnarray}
 S_{\rm Pb}& & (\WgPb=18.2\ {\rm GeV}) \nonumber \\
 && = 0.74\pm 0.07 \ ({\rm stat.}) \pm 0.07 \ ({\rm syst.}), 
 \label{eq:S1} 
 \end{eqnarray}

\begin{equation}
  S_{\rm Pb}(\WgPb=92.4\ {\rm GeV})= 0.62^{+0.04}_{-0.03} \ ({\rm stat.+syst.}),
  \label{eq:S2} 
  \end{equation}
  
\begin{eqnarray}
 S_{\rm Pb}& & (\WgPb=469.5\ {\rm GeV}) \nonumber \\
 & & =  0.47\pm 0.09 \ ({\rm stat.})\ ^{+0.06}_{-0.07} \ ({\rm syst.}). 
 \label{eq:S3} 
\end{eqnarray}

\begin{figure}
\includegraphics[width=0.48\textwidth]{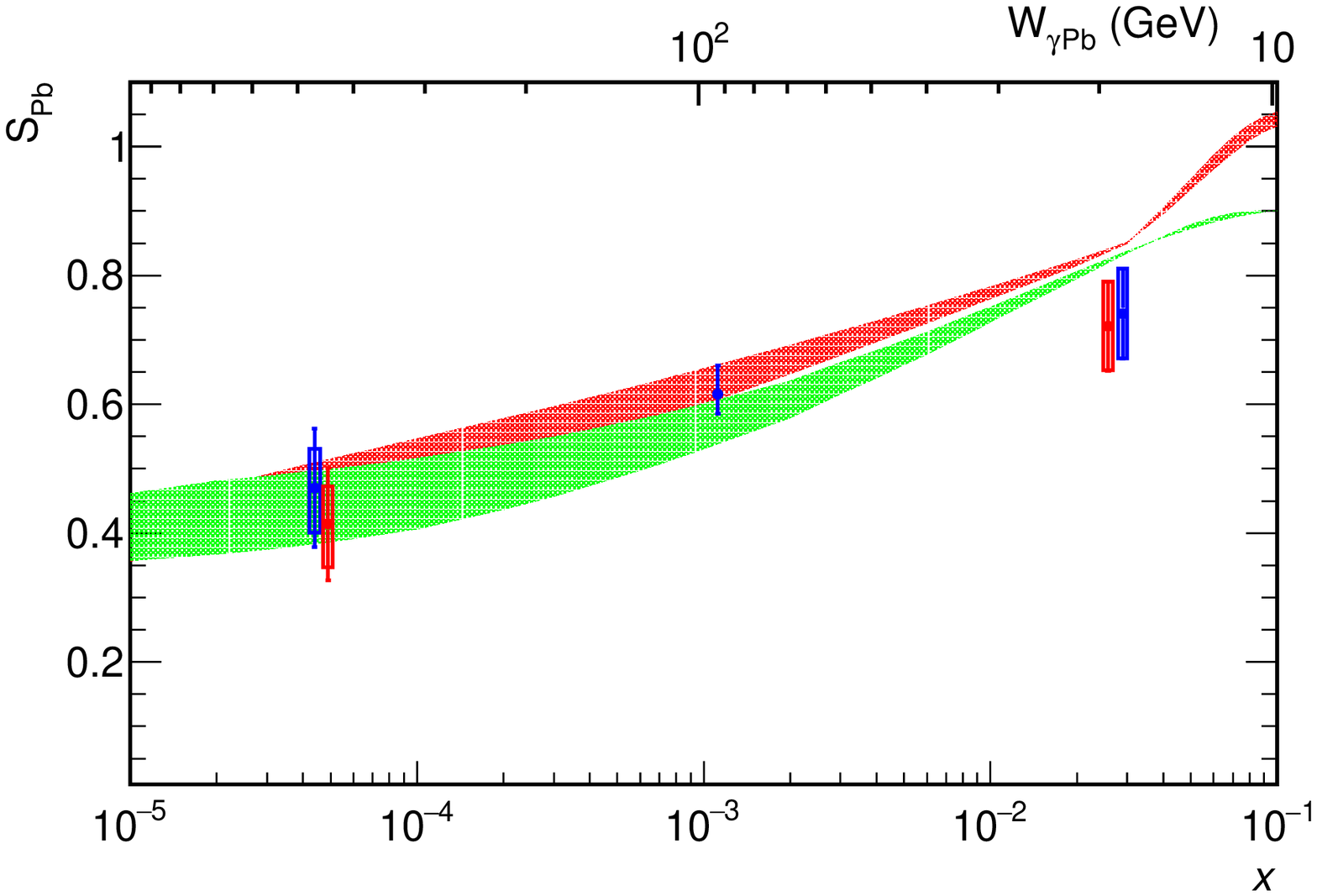}
\caption{\label{fig:Spb} Nuclear suppression factor  for $\jpsi$ coherent photoproduction off a lead nucleus 
as a function of $\WgPb$ (upper axis) and $x$ (lower axis). Statistical uncertainties are represented by a line and systematic uncertainties by the height of the empty box, except for the middle cross section, where the bars represent both types of uncertainties. Blue symbols are evaluated at the center, while red symbols are evaluated at the mean rapidity of the rapidity ranges where the corresponding measurements were performed.
The red (green) band correspond to the prediction of the LTA approach when using CTEQ6L (MNRT07) parton parameterizations. For details see~\cite{Guzey:2013qza}.}
\end{figure}

Figure~\ref{fig:Spb} shows these nuclear suppression factors as a function of both, $\WgPb$ and $x$.
In the absence of  nuclear effects, $S_{\rm Pb}(\WgPb)$ equals one. The fact that it is below one, indicates a suppression of the cross section off nuclear targets with respect to expectations from a  superposition of cross sections off free nucleons. This is direct evidence of nuclear shadowing. The facts that ($i$) at these values of $x$ both the nucleon and nucleus structure is dominated by gluons and ($ii$) this process involves at least two gluons,  suggest that (a large) part of this suppression is produced by  gluon shadowing. The same figure shows the results when using the mean rapidity value, instead of the center, of the region cover by the measurement. Again, although there are slight numerical changes, the overall picture is the same.

\subsection{Comparison to the leading twist approximation}

Nonetheless, the extraction of gluon shadowing will depend on each specific calculation. Gluons  in hadrons can be described with different types of distributions; e.g. collinear, unintegrated, generalized. In each of these cases the quantitative details will depend on the order of the calculation (LO, NLO, ...) and on the applied factorization and renormalization schemes.  

As an example of the potential impact of the cross sections given in (\ref{eq:XS1}) to (\ref{eq:XS3}) for computations of gluon shadowing, Fig.~\ref{fig:Spb} compares the nuclear suppression factors given in (\ref{eq:S1}) to (\ref{eq:S3}) to those obtained with the leading twist approximation (LTA) computed in~\cite{Guzey:2013qza}. The LTA calculation used  collinear parton distributions computed at leading order at a scale $\mu^2=3$ GeV$^2$. 
In this approach,   according to Eqs. (2.8) and (2.11) 
in~\cite{Guzey:2013qza}, the contribution to nuclear shadowing coming from gluon shadowing, is between 87\% to 97\% of $S_{\rm Pb}(\WgPb)$.  Figure~\ref{fig:Spb} shows that in LTA gluon shadowing reaches values $\lesssim$ 60\%
at $x\approx 5$x10$^{-5}$.

\section{Discussion\label{sec:discussion}}

Some comments are in order. As mentioned in the Introduction, the interpretation of the excess measured in \cite{Adam:2015gba} as due to coherent $\jpsi$ photoproduction is based on the "{\it plausible assumption}" that this is the right explanation. The assumption is quite reasonable, so it has been taken as correct in this work.

Up to now it is not clear what is the meaning of coherence in peripheral collisions. To my knowledge there is currently only one  study addressing this issue~\cite{Klusek-Gawenda:2015hja}. Taking into account that in the 70\%--90\% centrality class the number of participants is just a few percent of the possible participants (see Table I in~\cite{Abelev:2013qoq}), 
it seems that considering a coherent interaction with the full nucleus is a reasonable approximation for the current size of experimental uncertainty. Note that a similar approach has been taken in the rough estimations presented in ~\cite{Adam:2015gba}, which produced cross sections around 40 $\mu$b; lower than what was measured, but still compatible with the measurement.  
Furthermore, the distribution of transverse momentum of the produced $\jpsi$ measured by ALICE (see Fig. 1 in~\cite{Adam:2015gba}) seems to be compatible, although slightly wider,  with the distribution of coherent $\jpsi$ obtained in UPC.

One of the advantages of the method presented here is that the measurement of coherent $\jpsi$ photoproduction in peripheral and ultra-peripheral collisions at $y=0$ would allow to test quantitatively how good is this approximation, by computing the fluxes in both scenarios with Eqs.~(\ref{eq:nU}) and~(\ref{eq:nP}) and plugging them in Eq.~(\ref{eq:XS}). The extracted $\sigma_{\gamma \rm Pb}$ in both cases, should be the same. Given that currently there is already one measurement, it could be used to predict that the Pb-Pb cross section measured at mid-rapidity in the 70\%--90\% class for Run 1 energies should be

\begin{eqnarray}
\frac{d\sigma^P_{\rm PbPb}}{dy} (y=0)& = & \frac{n^P_\gamma(y=0)}{n^U_\gamma(y=0)}   \frac{d\sigma^U_{\rm PbPb}}{dy}(y=0) \nonumber \\
%& = & \frac{5.2}{66.5} \left\{ 2.38^{+0.34}_{-0.24}\ {\rm (stat.+syst.)} \ {\rm  mb} \right\} \nonumber \\
& = & 186^{+27}_{-19}\ {\rm (stat.+syst.)} \ {\rm \mu b}.
\label{eq:pred}
\end{eqnarray}

The experimental uncertainties from Run 1 measurements are of a few tens of persent, which justify that uncertainties in the photon flux or the computation of the impulse approximation have been neglected. These uncertainties have been studied in~\cite{Guzey:2013xba} and estimated to be at most at the level of a few percent. If one would add in quadrature these few percent contributions, to the much larger experimental uncertainties, the total uncertainty would  not change. Measurements from Run 2 are expected to be more precise and a more careful treatment of experimental and theoretical uncertainties would be needed.

It would also be important if the CMS Collaboration could measure the peripheral process at the same rapidity as their current UPC measurements~\cite{Khachatryan:2016qhq}. This would allow to over-constraint the system of Eqs.~(\ref{eq:SigYneg}) to (\ref{eq:Flux}) and would improve the extraction of $\sigma_{\gPb} (\WgPb)$.

In the same spirit, it would be important to repeat these measurements with Run 2 data. The factor of two increase in $\sNN$, means that larger (smaller) values of $\WgPb$ ($x$) can be reached.  It would be very important that the measurements in the peripheral and the UPC classes are performed at exactly the same rapidity, to avoid the model dependence implicit in the shifting of UPC cross sections. It would also be important to perform each of  these measurements in a rapidity range as small as experimental considerations would allow.

Reference~\cite{Guzey:2013xba} appeared some years before the measurements  in~\cite{Adam:2015gba} were published. As such, those authors could not applied the method presented here. None the less, they attempted a first extraction of  $\sigma_{\gamma \rm Pb}$. For this, they made the approximation that the high-energy contribution to Eq.~(\ref{eq:XS}) could be neglected. In this case, they were able to extract the low-energy contribution by computing the photon flux. The results reported in~\cite{Guzey:2013xba} for the 
low-energy contribution and for the cross section at mid rapidity agree with those reported here in Eqs.~(\ref{eq:XS1}) and (\ref{eq:XS2}), respectivley. This reflects, that the approximation used by the authors of~\cite{Guzey:2013xba} was appropriate in the phase space that was used. But, by making this approximation, they could not obtain the result reported here in Eq.~(\ref{eq:XS3}).

There is another proposal based on accessing different fluxes at the same rapidity to disentangle both contributions in Eq.~(\ref{eq:XS}). It exploits  the fact that independent  electromagnetic interactions between the outgoing lead nuclei may excite them and produce forward neutrons. This process modifies the impact parameter dependence of the photon flux~\cite{Baltz:2002pp}.  Separating the events from UPC in classes defined by the number of measured forward neutrons allows to separate the $+y$ and $-y$ contributions~\cite{Guzey:2016piu}. 

The proposal presented here and that in~\cite{Guzey:2016piu} are complementary in the sense that they use different processes to extract the same cross section, so they are subjected to different experimental and theoretical uncertainties.  Up to now, there are no experimental results using this alternative approach. Once they become available, the comparison of the results in these two approaches will help to understand better the extraction of $\sigma_{\gamma \rm Pb}$ and of nuclear shadowing from data on coherent $\jpsi$ photoproduction in Pb-Pb collisions.

\section{Summary and outlook\label{sec:so}}

In summary,  a novel method to disentangle the two contributions to the coherent photoproduction of $\jpsi$ in Pb-Pb collisions and thus to extract $\sigma_{\gamma \rm Pb}$  has been presented. 
The application of the method has been illustrated using current existing data from ALICE, obtaining   
$\sigma_{\gamma \rm Pb}$ up to a photon-lead centre-of-mass energy of  470 GeV.
In order to bring these results closer to the concept of nuclear shadowing the nuclear suppression factor for this process has been computed. The results have been compared to a prediction based on the LTA approximation. 

Future data from Run 2 at the LHC will allow to quantify the precision of this method using data at mid rapidity and will constrain better the extracted cross section adding more data points to the procedure.  Furthermore, data from Run 2 will allow to reach larger (smaller) values of $\WgPb$ ($x$). The availability of a precise nuclear suppression factor for this process, which is mediated by at least two gluons, and in such small-$x$ regime, will advance the understanding of gluon shadowing putting strong constraints to theoretical predictions of this phenomenon.

\section*{Acknowledgements}
The author would like to thank G. Martinez for  early discussions on this method. I would also like to thank V. Guzey for providing the LTA predictions shown in Fig.~\ref{fig:Spb}. This work was partially supported by grant LK11209 of  M\v{S}MT \v{C}R.

\end{document}